\documentclass[reprint,twocolumn,floats, floatfix,superscriptaddress,amsmath,amssymb,aps]{revtex4-1}

\usepackage{graphicx}
\usepackage{dcolumn}
\usepackage{bm}
\usepackage{color}
\usepackage{lipsum}
\usepackage[normalem]{ulem}  
\usepackage{xcolor}

\begin{document}

\title{Reentrant Rigidity Transition in Planar Epithelia with Volume- and Area Elasticity}

\author{Tanmoy Sarkar}
\affiliation{Jo\v zef Stefan Institute, Jamova 39, SI-1000 Ljubljana, Slovenia}%

\author{Matej Krajnc}
\affiliation{Jo\v zef Stefan Institute, Jamova 39, SI-1000 Ljubljana, Slovenia}

\begin{abstract}
{We find a reentrant columnar-to-squamous rigidity transition in 3D epithelia, governed by volume- and area elasticity. Our framework maps onto the classic 2D Area- and Perimeter-Elasticity model but, unlike its 2D counterpart, exhibits both softening and stiffening depending on the initial state. The phase diagram reveals floppy states with vanishing shear and in-plane bulk moduli, alongside a lateral-tension–driven discontinuous columnar-to-squamous transition. The critical behavior underlying the emergence of the reentrant transition belongs to the mean-field universality class.}
\end{abstract}

\maketitle

{\it Introduction.---}During morphogenesis, epithelial tissues must have an ability to transition dynamically between elastic and fluid-like states to support mechanical stresses on one hand and facilitate cellular rearrangements and large-scale tissue flows on the other~\cite{mongera18,tetley19,kim20,atia21,lawsonkeister21,petridou21,hannezo22}. This dual solid–fluid behavior has been widely studied using two-dimensional~(2D) Area- and Perimeter-Elasticity~(APE) model, which assumes that each cell's apical area $A$ and perimeter $P$ tend towards preferred values $A_0$ and $P_0$, respectively~\cite{farhadifar07,Hocevar09,staple10,bi15,park15}. The model predicts a rigidity transition controlled by the preferred cell shape: Tissues are rigid when the actual cell shapes are incompatible with the preferred one and floppy when cells can attain their preferred areas and perimeters~\cite{moshe18,merkel19}. In the floppy regime, the linear shear modulus vanishes and, depending on their structure, tissues may also display zero energy barriers for cell rearrangements~\cite{bi15,sahu20}.

These findings extend to fully three-dimensional (3D) space-filling cell aggregates~\cite{merkel2018geometrically}. Analogous to 2D, the generalized 3D model employs Volume- and Area Elasticity~(VAE), minimizing the energy when cell volumes and surface areas match the preferred values $V_0$ and $S_0$, respectively. As in 2D, the preferred 3D cell shape controls rigidity and the linear shear modulus vanishes when tissues transition from rigid to floppy states~\cite{merkel2018geometrically}. 

Generalizing the 2D APE model to 3D epithelial monolayers has proven more subtle. A natural approach is to couple the APE framework—capturing the mechanics of apically positioned adherens junctions and actomyosin cortices—to the basolateral cell surfaces under surface tension. Yet this apico–basolateral coupling suppresses the rigidity transition, rendering tissues rigid regardless of the preferred apical cell shape~\cite{rozman24}. This raises the central question of our work: Do the physics underlying the 2D APE model carry over to 3D monolayers? More specifically, what theoretical framework can account for a rigidity transition in 3D monolayers in a way that remains consistent with the 2D APE model?

Here, we address these questions by studying the mechanics of 3D monolayers governed by VAE. By 2D-projecting our model, we explore its relation to APE model and confirm that the properties of epithelia, studied in 2D, are generally translatable to 3D. We characterize, both analytically and numerically, the phase diagram of rigid and floppy monolayers and find a novel reentrant rigidity transition, controlled by in-plane isotropic strain. The model also features a regime of floppy states with a vanishing 2D bulk modulus, associated with isotropic in-plane deformations, and a discontinuous Columnar-to-Squamous Transition~(CST), controlled by cell-cell adhesion. Importantly, our study shows that 3D cell shapes fundamentally affect tissue's response to isotropic compression. While the 2D model, counterintuitively, predicts tissue softening upon compression and stiffening upon dilation~\cite{tong22,hernandez23}, tissues in our 3D model are rigidified both when compressed and dilated. 

\begin{figure}[b!]
    \includegraphics[]{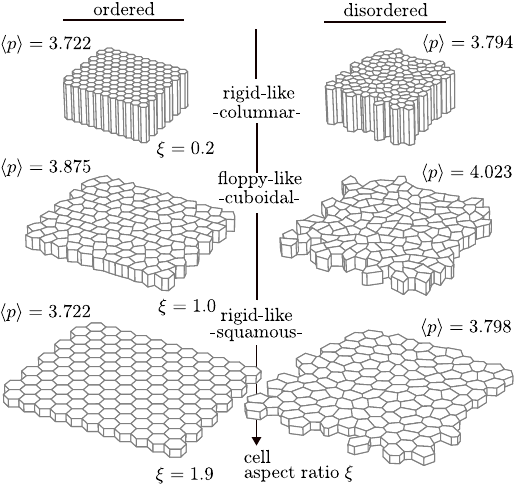}
    \caption{Simulation snapshots of CST. Ordered~(left) and disordered~(right) tissues for $S_0=5.75$ and $5.9$, respectively, and $\gamma=0.5$. For clarity, $\sim 10\%$ of all cells are shown. The fraction of non-hexagons in disordered sample $d=0.437$~\cite{supplMat}.}
    \label{F1}
\end{figure}

{\it The model.---}We consider ordered and disordered planar packings of $N_c=1024$ prismatic cells with periodic boundary conditions, assuming uniform cell height across the tissue~(Fig.~\ref{F1} and~\footnote{The assumption of uniform cell height is reasonable because cells in a monolayer are tightly packed. If apical vertices were free to move in the $z$-direction, the apical surface of the tissue would no longer remain generally flat. Nevertheless, the simplification of flat apical surface does not affect the main conclusions of our work}).
In addition to VAE that penalizes deviations of cell volumes $V_k$ and cell surface areas $S_k$ from preferred values, our potential energy includes a constant contribution to the lateral surface tension, $-\gamma$, describing cell-cell adhesion at the lateral cell sides; we denote the lateral cell surface areas by $A_k^{(l)}$ and, for generality, we consider both positive and negative $\gamma$-values. In dimensionless form, where units of energy and length are given by $k_SV_0^{4/3}$ ($k_S$ being the area-elasticity modulus) and $V_0^{1/3}$, respectively, the potential energy reads
\begin{equation}
    \label{eq:W}
    W=\sum_k\left [k_V\left (V_k-1\right )^2+\left (S_k-S_0\right )^2-\frac{\gamma}{2}A^{(l)}_k\right ]\>.
\end{equation}
Here $k_V$ is a dimensionless cell-incompressibility modulus; for simplicity, we begin by treating cells as nearly incompressible, i.e., $k_V=100$, and we show in Supplemental Material, Sec.~IV~\cite{supplMat} that finite $k_V$ does not fundamentally affect the main results of our work. Details related to the implementation of the vertex model and statistical properties of disordered cell packings are given in Supplemental Material, Sec.~I and II and Fig. S1~\cite{supplMat}.

We simulate CST in both ordered and disordered cell monolayers by quasistatically and isotropically increasing the simulation-box size~(Fig.~\ref{F1}). Due to cell incompressibility, the in-plane area increase implies an increase of the 3D cell aspect ratio~(i.e., width-to-height ratio), defined as
\begin{equation}
    \label{eq:xi}
    \xi=\frac{\sqrt{A}}{h}\overset{V=1}{=}A^{3/2}\>,
\end{equation} 
where $A$ and $h=A^{-1}$ are the in-plane cell area and height, respectively. Our simulations assume apico-basal symmetry in cell shapes, however, even in the absence of this constraint, cell shapes {\it are} apico-basally symmetric, consistently with the inherent apico-basal symmetry of the mechanics~[Eq.~(\ref{eq:W})]. The exception are highly columnar disordered tissues, where cells with pentagonal bases become either apically or basally constricted~\cite{krajnc18,sui18}. Importantly, this instability~\footnote{Apical and basal polygon collapses are known from other 3D monolayer models and are typically prevented by auxiliary energy terms that penalize vanishing polygon areas~\cite{sui18}.} does not break the global apico-basal symmetry of the tissue, in contrast to the asymmetry observed in the model by Rozman {\it et al.}~\cite{rozman24}.

We measure average in-plane cell-shape indices, defined as $\langle p\rangle=\langle P_k/\sqrt{A}\rangle$, where $P_k$ is the cell in-plane perimeter, and find that $\langle p\rangle$ non-monotonically depends on $\xi$, such that the in-plane structure of columnar and squamous tissues appears rigid-like with cells assuming highly isometric, low-$\langle p\rangle$, shapes, whereas cuboidal tissues appears floppy-like with high-$\langle p\rangle$ cell shapes~(Fig.~\ref{F1}).
\begin{figure*}[htb!]
    \includegraphics[]{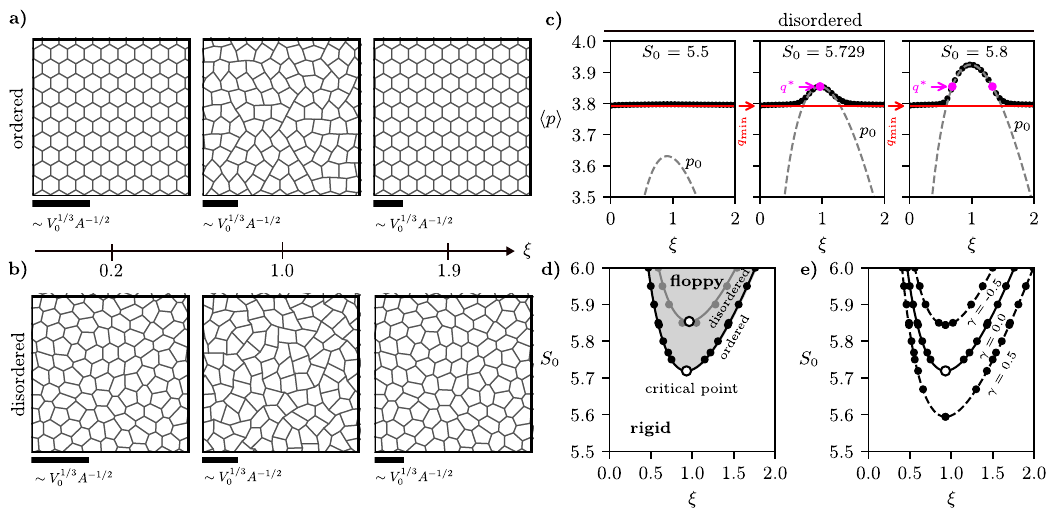}
    \caption{(a,b)~Simulation snapshots of CST transition of ordered tissue for $S_0=5.75$~(a) and disordered tissue for $S_0=5.9$~(b). In both cases $\gamma=0.5$. 3D cell aspect ratio $\xi$ increases from left to right. Scale bars indicate the absolute length scale $\sim V_0^{1/3}A^{-1/2}$. For clarity, $\sim 10\%$ of cells are shown. (c)~Average renormalized in-plane cell perimeter $\langle p\rangle$ versus $\xi$ in disordered tissue with fraction of non-hexagons $d=0.437$ for $S_0=5.5,\>5.729,$ and $5.8$ (datapoints); $\gamma=0.5$. Dashed gray curve and solid red line show $p_0$~[Eq.~(\ref{eq:p0})] and $q_{\rm min}$, respectively. Tissues are compatible for $\langle p\rangle>q^*$; here $q^*$ is the lowest $p_0$-value where all cells are compatible with the preferred shape. (d,e)~Rigidity diagram $S_0(\xi, \gamma)$ for $\gamma=0$ (d) and for ordered tissues at $\gamma=-0.5,\>0,$ and $0.5$~(e). Black and gray datapoints show simulation data and hollow circles indicate critical points. Phase boundaries [Eq.~(\ref{eq:phaseBoundary})] are shown by black and grey curves for ordered and disordered ($d=0.437$) tissues, respectively~(d). In simulations, tissues are classified as being compatible (floppy) when $\left |\langle p\rangle-p_0\right |$ is below a threshold. Because $\left |\langle p\rangle-p_0\right |$ decreases sharply at the transition, the classification is insensitive to the precise value of this threshold.}
    \label{F2}
\end{figure*}

{\it Mapping to the 2D APE model.---}The results of our 3D simulations~(Fig.~\ref{F1}) are indicative of a reentrant columnar-to-squamous rigidity transition. To assess their possible relation with the well-studied 2D APE model, we next project our model to 2D. In line with our vertex model, where $k_V=100$, we assume $k_V\to\infty$ and thus, $V_k=1$ for all cells. By expressing cell geometry in terms of $A$, $h$, and $P_k$ and accounting for the fixed-volume constraint, the single-cell energy becomes 
\begin{equation}
    W_k=A^{-2}\left [P_k-A\left (S_0-2A\right )\right ]^2-\frac{\gamma P_k}{2A}\>.
\end{equation}
Since, at a given 3D cell aspect ratio $\xi$, the cell in-plane area is {\it a priori} imposed~(Eq.~[\ref{eq:xi}]), $A$ can be used to renormalize cell perimeters $P_k$ as well as the energy. The single-cell energy, recast in terms of renormalized perimeter $p_k=P_k/\sqrt{A}$, and rescaled by $A^{-1}$, reads
\begin{equation}
    \label{eq:wk_renor}
    w_k=\left (p_k-p_0\right )^2+w_0\>,
\end{equation}
where
\begin{equation}
    \label{eq:p0}
    p_0=A^{1/2}\left (S_0-2A+\frac{\gamma}{4}\right )
\end{equation}
is the preferred renormalized cell perimeter and $w_0=\gamma A\left (16A-8S_0-\gamma\right )/16$. This mapping shows that, for any given in-plane cell area, set by in-plane strain, our $k_V \to \infty$-model maps exactly to a model describing a planar packing of unit-area polygons whose perimeters $p_k$ tend toward $p_0$. In terms of $p_k$, $w_0$ can be seen merely as an energy offset.

{\it Reentrant rigidity transition.---}This mapping allows us to simulate CST of 3D monolayers in 2D by minimizing the renormalized energy [Eq.~(\ref{eq:wk_renor})], while keeping all in-plane polygon areas fixed at 1. Due to the renormalization of cell sizes, the simulation-box size is kept fixed as well~(Figs.~\ref{F2}a and b), but the in-plane strain and thus the change of the 3D cell aspect ratio $\xi$, associated with CST, is effectively simulated by quasistatically varying $p_0$ (and $w_0$) according to Eq.~(\ref{eq:p0}).

At a given cell aspect ratio $\xi$, the minimal possible renormalized energy $w=N_c w_0$~[Eq.~(\ref{eq:wk_renor})] is reached if all cells assume the preferred renormalized perimeter $p_0$. Such states are not possible if $p_0$ is too small, as every packing of equal-area polygons is characterized by a minimal attainable average cell-shape index $q_{\rm min}$. This minimal value depends on the packing's statistical properties: For a regular hexagonal packing, $q_{\rm min}=q_6=2^{3/2}\times 3^{1/4}\approx3.722$, corresponding to the perimeter of unit-area regular hexagon, whereas in disordered packings, $q_{\rm min}>q_6$~(Supplemental Material, Fig.~S2a,b~\cite{supplMat}). Tissues with $p_0<q_{\rm min}$, therefore, necessarily describe geometrically incompatible, i.e., rigid, states. Plotting average renormalized in-plane cell perimeters $\langle p\rangle$ versus $\xi$ shows that for small enough $S_0$-values, these incompatible states appear along the whole range of 3D cell aspect ratios~(Fig.~\ref{F2}c, left, and Supplemental Material, Fig.~S2d~\cite{supplMat}) and $\langle p\rangle$ in this regime is close to $q_{\rm min}$.

Geometrically compatible, i.e., floppy, states are present only for large enough $S_0$-values and appear for 3D cell aspect ratios close to unity, i.e., cuboidal tissues~(Fig.~\ref{F2}c, middle and right, and Supplemental Material, Fig.~S2d~\cite{supplMat}), as observed in 3D simulations of CST~(Fig.~\ref{F1}). Compatibility requires all cells assuming the preferred in-plane area~(Supplemental Material, Fig.~S3~\cite{supplMat}) and the preferred in-plane perimeter. Since renormalized in-plane cell perimeters of all cells in regular-hexagonal packings can attain the minimal cell-shape index ($p_k=q_{\rm min}=q_6$), $q_6$ is at the same time the lowest $p_0$-value where the compatibility condition is met; we denote this value more generally by $q^*$. In contrast, in disordered tissues where perimeters are distributed, the lowest $p_0$-value where {\it all} cells are compatible with the preferred shape $p_0$ is $q^*>q_{\rm min}$. Like $q_{\rm min}$, $q^*$ too depends on the degree of disorder~(Supplemental Material, Fig.~S2b,c~\cite{supplMat}). Thus, the general compatibility condition $p_k=p_0=q^*$ yields the rigid-floppy boundary in the $S_0(\xi)$ diagram:
\begin{equation}
    \label{eq:phaseBoundary}
    S_0(\xi)=\frac{q^*+2\xi}{\xi^{1/3}}-\frac{\gamma}{4}\>.
\end{equation}
This relation agrees perfectly with simulations~(Fig.~2d,e and Supplemental Material, Fig.~S2e,f~\cite{supplMat}) and describes a reentrant rigidity transition, controlled by the in-plane isotropic strain through $\xi$.

The reentrant transition appears only for sufficiently high $S_0$: At $\xi^*=q^*/4$, where $\partial_\xi S_0=0$, the phase diagram features a critical point $S_0^*=3\cdot 2^{-1/3}q^{*2/3}$. For $S_0<S_0^*$, there are no floppy states, regardless of the $\xi$-value. For hexagonal cell packing, $\xi^*=2^{-1/2}\times 3^{1/4}\approx 0.93$ and $S_0^*=3^{7/6}\times 2^{2/3}\approx 5.719$.

An intuitive explanation for the observed reentrant rigidity transition is as follows: As the 3D cell aspect ratio deviates from a compatible cuboidal shape--either by becoming columnar or squamous--the total cell surface area increases. To maintain compatibility, this increase may be compensated by reducing the lateral surface area. Since for a given cell aspect ratio, cell height as well as apical and basal areas are fixed, the lateral surface area can only be reduced by shortening of the in-plane cell perimeter. However, when the average cell-shape index reaches $q^*$, this perimeter adjustment is no longer possible and tissue rigidifies.

\begin{figure}[t!]
    \includegraphics[]{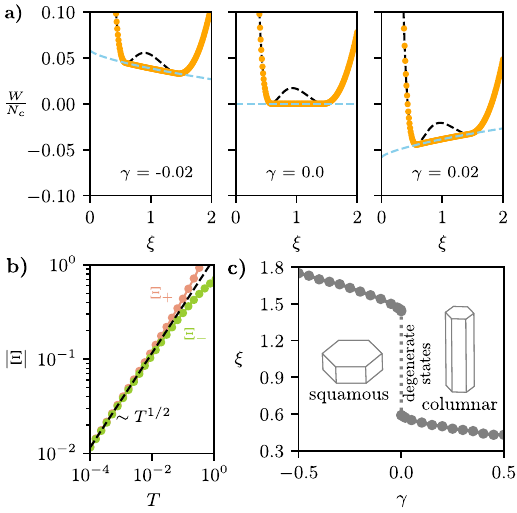}
    \caption{(a)~Total energy per cell, $W/N_c$, as a function of $\xi$ for $\gamma=-0.02,\>0,$ and $0.02$ (from left to right). Orange datapoints show numerical results, whereas dashed blue and black curves show $W_{\rm floppy}=W(P=p_0\sqrt{A})$ and $W_{\rm rigid}=W(P=q_6\sqrt{A})$, respectively. (b) The $T^{1/2}$-critical scaling of the ``order parameter'' $\left |\Xi\right |$. The two branches $\Xi_\pm$ denote the reentrant transition points $\xi_\pm-\xi^*$. (c)~Equilibrium cell aspect ratio $\xi$ versus lateral tension $\gamma$. In panels (a) and (c), a regular-hexagonal packing with $S_0=5.85$ is considered.}
    \label{F3}
\end{figure}

{\it Mean-field theory.---}To examine the emergence of the rigidity transition as $S_0$ crosses the critical point, we study the energy of a regular hexagonal cell arrangement, $F=W\left (P=q_6\sqrt{A}\right )$, recast in terms of the ``order parameter" $\Xi=\xi-\xi^*$ and generalized ``temperature" $T=S_0-S_0^*$~(Supplemental Material, Sec.~III~\cite{supplMat}). We perform a 4-th order Taylor expansion of $F$ which for $\gamma=0$ reads
\begin{equation}
    \label{meanfieldF}
    F\approx F_0+a_2(T)\Xi^2+a_3(T)\Xi^3+a_4(T)\Xi^4\>.
\end{equation}
Here, the quadratic coefficient scales linearly with $T$, as in the classical mean-field theory of the Ising model. The difference lies in the presence of the cubic term, whose presence at first glance suggests a discontinuous transition similar to the Landau-de Gennes theory of nematic-isotropic phase transition in liquid crystals. However, since $a_3$, like $a_2$, vanishes linearly with $T$, while the stabilizing quartic term approaches a positive constant~(Supplemental Material, Sec.~III~\cite{supplMat}), the resulting critical behavior coincides with that of the classical Ising mean-field theory with the $T^{1/2}$-critical scaling of the ``order parameter", $\left | \Xi\right |=3^{2/3}2^{-5/6}T^{1/2}$, describing the emergence of the reentrant transition points $\xi_-$ and $\xi_+$ as $T$ becomes positive~(Fig.~\ref{F3}b).

For aspect ratios between $\xi_-$ and $\xi_+$~(Fig.~\ref{F3}a, middle), where cells are compatible with the preferred shape, the energy landscape, described by $W_{\rm floppy}=W\left (P=p_0\sqrt{A}\right )$, is flat, indicating that, in addition to the vanishing shear modulus, floppy states are also characterized by a zero in-plane bulk modulus, associated with isotropic in-plane deformations.
Unlike in 2D, such deformations may occur at a constant energy since (i)~3D cell volumes can be preserved under in-plane strain by adjusting cell height, whereas (ii) the change of the cell surface area, associated with varying cell aspect ratio, may be prevented by adjusting the in-plane polygon shape factor as discussed above. Notably, in 2D, similar energy-free isotropic deformations exist only in the trivial case, where cell-area elasticity is absent~\cite{merkel19,hernandez23}. We note that resistance to cell-height fluctuations that may arise from the elasticity of cytoplasmic components~\cite{hannezo14} may lift this energy degeneracy~(Supplemental Material, Fig.~S4~\cite{supplMat}).

Expanding the analogy with the classical mean-field Ising model, we observe that the lateral tension $\gamma$ contributes to the energy $F$ [Eq.~(\ref{meanfieldF})] through an additional term $h\Xi$, where $h=2^{7/6}\cdot 3^{-13/12}\gamma$ is a generalized external field. This field breaks the “up–down” symmetry, favoring squamous cells with $\xi=\xi_+$ when $\gamma < 0$ and columnar cells with $\xi=\xi_-$ when $\gamma > 0$~(Fig.~\ref{F3}a and c). Near the critical point, the optimal cell aspect ratio $\xi = \xi^* + \chi \gamma$, with a generalized susceptibility $\chi \propto |T|^{-1}$, satisfying the universal ratio $\chi(T<0)/\chi(T>0) = 2$. At $T=0$, the relation between $\xi$ and $\gamma$ is given by the critical ``isotherm" $\gamma \propto -\Xi^3$. Altogether, these critical scaling exponents are consistent with Landau mean-field universality class~(Supplemental Material, Sec.~III~\cite{supplMat}).

{\it Discussion.---}Our results show that the rigidity of epithelia with VAE~[Eq.~(\ref{eq:W})] may be tuned non-monotonically by in-plane isotropic strain~(Fig.~1), describing a reentrant CST rigidity transition, whereby tissues transition from rigid columnar- to rigid squamous states, through intermediate floppy cuboidal states~(Fig.~2). An additional constant lateral tension, distinguishing lateral from apical and basal sides, introduces a discontinuous columnar-squamous transition with energy-degenerate states with zero in-plane bulk modulus, when lateral tension is absent~(Fig.~3).

Interestingly, our 2D-projected model shows that varying 3D cell aspect ratio, i.e., by imposing in-plane isotropic strain, effectively corresponds to nonlinearly changing the effective in-plane preferred cell-shape index $p_0$~[Eq.~(\ref{eq:p0})], which in turn controls the rigidity transition. This provides an alternative physical interpretation of $p_0$, which has been so far interpreted through a competition between cortical tension and adhesion strength~\cite{farhadifar07,bi15}. 

2D APE model, which inherently does not describe cell height, counterintuitively and in contrast with the standard jamming~\cite{liu98}, predicts tissue softening upon in-plane compression~\cite{tong22,hernandez23}. By accounting for the third dimension, our model changes this view. Indeed, compressing and dilating a monolayer away from a cuboidal state both lead to its rigidification~(Fig.~\ref{F1}).

Our analysis also shows that VAE monolayer model is a more direct 3D analog of the 2D APE model than the model in which apical cell sides, described by APE, are coupled to the basolateral cell domains under surface tension~\cite{rozman24}. Indeed, while the apico-basolateral coupling in the latter model prevents the rigidity transition altoghether, our 3D VAE monolayer model preserves it and in the limit of incompressible cells even exactly maps to 2D fixed-area APE model. We note that the rigidity transition persists even in a generalized version of our model, where cells are associated both with the preferred surface area and preferred apical perimeter~(Supplemental Material, Fig.~S5~\cite{supplMat}).

Finally, while the assumption that epithelial cells tend to preserve both their volumes and total surface areas is yet to be experimentally validated, this phenomenology may readily apply to studying synthetic planar packings of strongly adhering lipid vesicles, where the membrane's bending elasticity may be negligible and the mechanics may be well-described by the VAE model~\cite{Hocevar11}.

{\it Acknowledgements.---}We thank Silke Henkes and members of her research group at the Lorentz Institute in Leiden for fruitful discussions, as well as Jan Rozman and Primo\v z Ziherl for suggesting a generalization of our initial model. We acknowledge the financial support from the Slovenian Research and Innovation Agency (research projects J1-3009 and J1-60013, development funding pillar RSF-0106, and research core funding P1-0055). 
\bibliography{apssamp}

\end{document}



\title{Reentrant Rigidity Transition in Planar Epithelia with Volume- and Area Elasticity:\\Supplemental material}

\author{Tanmoy Sarkar}
\affiliation{Jo\v zef Stefan Institute, Jamova 39, SI-1000 Ljubljana, Slovenia}%

\author{Matej Krajnc}
\affiliation{Jo\v zef Stefan Institute, Jamova 39, SI-1000 Ljubljana, Slovenia}%

\maketitle
\onecolumngrid

\section{3D monolayer vertex model}
The total potential energy of the cell monolayer in dimensionless form reads
\begin{equation}
    \label{eq:W}
    W=\sum_k\left [k_V\left (V_k-1\right )^2+\left (S_k-S_0\right )^2-\frac{\gamma}{2}A^{(l)}_k\right ]\>,
\end{equation}
where the sum goes over all cells $k$. Parameters $k_V$, $S_0$, and $\gamma$ are the cell-compressibility modulus, the preferred cell surface area, and cell-cell adhesion strength, respectively. The geometric variables $V_k$, $S_k$, and $A_k^{(l)}$ are cell volume, cell surface area, and cell lateral surface area, respectively.

The energy [Eq.~(\ref{eq:W})] is minimized, using gradient descent, where degrees of freedom, i.e., vertex positions, $\boldsymbol r_i=\left (x_i,y_i,z_i\right )$, are propagated by $\eta\dot{\boldsymbol r}_i=-\nabla_i W$. The integration is performed using an explicit Euler's integration scheme:
\begin{equation}
    \boldsymbol r_i(t+\Delta t)=\boldsymbol r_i(t)-\frac{\Delta t}{\eta}\nabla_iW(\boldsymbol r)\>,
\end{equation}
with $\Delta t/\eta=10^{-3}$. The energy gradient $\nabla_i W$ is calculated as
\begin{equation}
\label{eq:grad_W}
	\nabla_i W=\sum_{k}\left (2\kappa_V(V_k-1)\nabla_iV_k+2(S_k-S_0)\sum_{m\in S_k}\nabla_i A_m-\frac{\gamma}{2}\sum_{m\in A_k^{(l)}}\nabla_iA_m\right )\>,
\end{equation}
which further requires calculating gradients of polygon surface areas $A_m$ and cell volumes $V_k$ as described below; the inner sums in Eq.~(\ref{eq:grad_W}) go over all cell polygons, constituting cell's surface area $S_k$ and over all lateral polygons, constituting cell's lateral surface area $A_l^{(k)}$.

The surface area of the $m$-th polygon is calculated as a sum of surface areas of triangular surface elements, $\big\lVert \boldsymbol a_{m,\mu} \big\rVert$, defined by pairs of consecutive polygon vertices $\boldsymbol r_\mu$ and $\boldsymbol r_{\mu+1}$, and the polygon's center of mass 
\begin{equation}
    \boldsymbol c_{m}=\frac{1}{n_{m}}\sum_{\mu\in \langle m\rangle }\boldsymbol r_\mu\>.
\end{equation} 
Here, Greek indices do not denote the real vertex identification numbers, but their sequential indices within individual polygons. The surface area of polygon $m$ reads
\begin{equation}
	A_{m}= \sum_{\mu\in \langle m\rangle} \frac{1}{2}\big\lVert \left (\boldsymbol r_\mu-\boldsymbol c_{m}\right )\times\left (\boldsymbol r_{\mu+1}-\boldsymbol c_{m}\right ) \big\rVert
\end{equation}
and its gradient
\begin{align}
    \nabla_i A_{m}&=\sum_{\mu\in \langle m\rangle} \frac{1}{2}\nabla_i\Big\lVert\left (\boldsymbol r_\mu-\boldsymbol c_{m}\right )\times\left (\boldsymbol r_{\mu+1}-\boldsymbol c_{m}\right )\Big\rVert=\\
    &=\sum_{\mu\in \langle m\rangle}\frac{\big[(\delta_{i\mu} - n_{m}^{-1})(\mathbf{r}_{\mu+1} - \mathbf{c}_{m}) + (n_{m}^{-1}- \delta_{i(\mu+1)})(\mathbf{r}_\mu - \mathbf{c}_{m}) \big] \times \boldsymbol a_{m,\mu}}{4\big\lVert \boldsymbol a_{m,\mu} \big\rVert}\>,
\end{align}
where $\delta_{ij}$ is the Kronecker delta, whereas $\boldsymbol a_{m,\mu}$ is the area vector of the $\mu$-th triangular surface element of polygon $m$; the size of $\boldsymbol a_{m,\mu}$ is the corresponding surface area and its orientation is normal to the surface element.

The volume of the $k$-th is calculated as a sum of volumes of tetrahedra, defined by triangular surface elements $(\boldsymbol r_\mu,\boldsymbol r_{\mu+1},\boldsymbol c_{m})$ and with the fourth vertex at the origin $(0,0,0)$, as
\begin{equation}
	V_k=\sum_{m\in \langle k\rangle}\sum_{\mu\in \langle m\rangle}\frac{1}{6}\boldsymbol c_{m}\cdot\left (\boldsymbol r_\mu\times\boldsymbol r_{\mu+1}\right )\>.
\end{equation}
Its gradient is calculated as
\begin{align}
       \nabla_i V_l&=\sum_{m\in \langle k\rangle}\sum_{\mu\in \langle m\rangle}\frac{1}{6}\nabla_i\left [\boldsymbol c_{m}\cdot\left (\boldsymbol r_\mu\times\boldsymbol r_{\mu+1}\right )\right ] \nonumber\\
        &=\sum_{m\in \langle k\rangle}\sum_{\mu\in \langle m\rangle}\frac{1}{6 n_{m}} (\mathbf{r}_\mu \times \mathbf{r}_{\mu+1}) + \frac{1}{6} (\delta_{i\mu} \mathbf{r}_{\mu+1} - \delta_{i(\mu+1)} \mathbf{r}_\mu) \times \mathbf{c}_{m}\>.
\end{align}

\section{Sample preparation}
Since cells in the monolayer are represented by apico-basally symmetric prisms, we prepare both 2D and 3D tissue samples using the 2D apical vertex model. For 3D simulations, we extend the generated 2D samples to the third dimension by copying the apical polygonal packing to the basal side.

In the 2D apical vertex model, cell apical sides are represented by polygons, packed within a rectangular simulation box. We begin by generating a regular hexagonal lattice of cells, filling the box. While for simulations of ordered cell packings, this is already the required sample, simulations of disordered packings require further steps of sample preparation. 

We generate disordered tissues from the initially regular-hexagonal polygonal tiling by 1) melting the crystal using an active-tension-fluctuations scheme, described by the Ornstein-Uhlenbeck model~\cite{curran17,krajnc20,krajnc21}, and 2) ``quenching'' the sample by decreasing the fluctuations magnitude linearly with time. The ``quench'' rate determines the final statistical properties of the disordered sample: If the quenching is slow, final states will be more ordered compared to a fast quench.

Specifically, vertex positions undergo dynamics described by:
\begin{align}
    \label{eq:rdott_generalized}
    \bm{\dot{r}}_i = -2 \sum_k [k_A \left (A_k - 1\right ) \nabla_i A_k + (P_k - P_0) \nabla_i P_k] -\sum_j \Delta\gamma_j(t) \nabla_i l_j\>,
\end{align}
where $A_k$ and $P_k$ are polygon areas and perimeters, respectively, whereas $l_j$ is the length of edge $j$. Gradients of polygon areas, perimeters, and edge lengths are calculated analogously to the 3D calculation~(Section~I).

The fluctuating line tension, acting on edge $j$, $\Delta\gamma(t)$, evolves according to the Ornstein-Uhlenback scheme as 
\begin{equation}
    \label{eq:fluct}
    \Delta\dot{\gamma}_{j}(t) = -\frac{1}{\tau_m}\Delta\gamma_{j} + \sqrt{\frac{2 \sigma^2}{\tau_m }}\xi_{j}\>. 
\end{equation}
Here, the noise obeys $\langle \xi_{j}(t) \rangle = 0$ and $\langle \xi_{j}(t) \xi_{l}(t')\rangle = \delta_{jl}\delta(t-t')$ with $\sigma^2$ being its long-time variance. For all sample preparations, we use the following model-parameter values: $k_A=100$, $P_0=3.72$, and $\tau_m=1$.

\begin{figure*}[htb!]
    \includegraphics[]{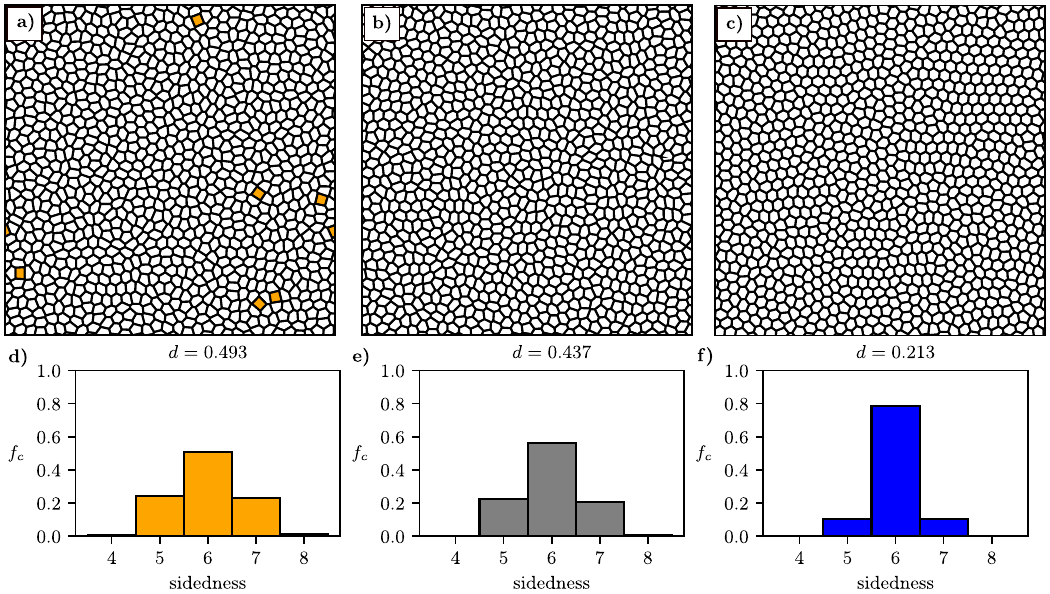}
    \caption{(a-c)~Snapshots of disordered samples, prepared by the procedure described in Sec.~II. Quadrilaterals (highlighted) are only present in the most disordered sample (a). (d-f)~Distributions of polygon sidedness for samples shown in panels a-c.}
    \label{FS1}
\end{figure*}

After initializing the sample as a regular hexagonal packing of cells, Eq.~(\ref{eq:rdott_generalized}) is simulated at $\sigma=0.5$ for 1000 time units. This is followed by the quench step during which $\sigma$ is linearly decreased with time from 0.5 to 0 as $\sigma(t)=0.5(1-r_qt)$; we considered quench rates $r_q=1/10$ and $1/500$~($d=0.437$ and $d=0.213$, respectively, in Fig.~\ref{FS1}) as well as an instantaneous step-like quench~($d=0.493$ in Fig.~\ref{FS1}). The final simulation step is performed at $\sigma=0$, where the system is allowed to dissipate the excess energy for another 1000 time units. During all 3 steps T1 transitions are allowed and performed on edges with lengths bellow a threshold $l_{\rm th}=0.01$. We distinguish between different disordered samples by 
\begin{equation}
    d=1-f_6\>,
\end{equation} 
where $f_6$ is the frequency of hexagons.

\begin{figure*}[htb!]
    \includegraphics[]{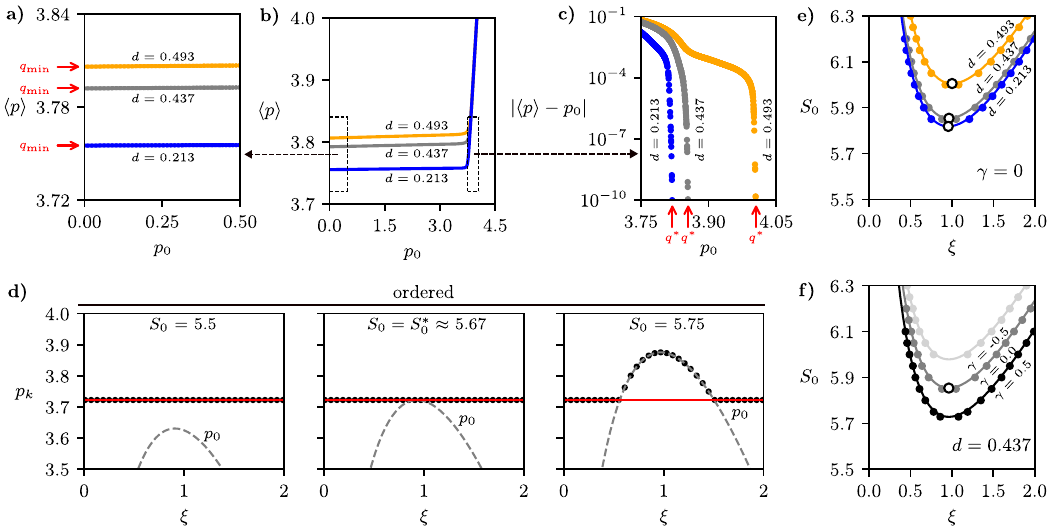}
    \caption{(a–c)~Average renormalized in-plane cell perimeter $\langle p \rangle$ vs. preferred perimeter $p_0$ in disordered tissues with varying disorder levels $d$. Note that $p_0$ depends nonlinearly on the cell aspect ratio $\xi$ as given by Eq.(5).
(a)~Zoom near small $p_0$ values; red arrows indicate minimum attainable $\langle p \rangle$, denoted by $q_{\mathrm{min}}$. For $d = 0.493$, $0.437$, and $0.213$, the corresponding $q_{\mathrm{min}}$ values are $3.806$, $3.792$, and $3.755$, respectively.
(b)~Full range of $p_0$ values; dashed boxes mark regions shown in (a) and (c).
(c)~Absolute deviation $|\langle p \rangle - p_0|$ vs. $p_0$; red arrows mark $q^*$ where $\langle p \rangle$ begins to match $p_0$ up to the precision $\sim 10^{-10}$. For $d = 0.493$, $0.437$, and $0.213$, the corresponding $q^*$ values are $4.005$, $3.855$, and $3.820$, respectively.
(d)~Renormalized cell perimeters versus cell aspect ratio $\xi$ for a regular hexagonal lattice [analogous to Fig. 2c from the main text]. (e, f)~Rigidity diagram $S_0(\xi, \gamma)$ of disordered tissue. Phase boundaries are shown in solid lines, while filled circles and hollow circles indicate simulation data points and critical points, respectively. (e) Effect of disorder on the phase boundary at $\gamma=0$. In panels c and e, the $d=0.493$-case deviates from the other two samples since it is the only sample that includes quadrilaterals, whose minimal shape index is 4. (f)~Influence of $\gamma$ on the phase boundary at a fixed disorder level $d=0.437$.}
    \label{FS2}
\end{figure*}

\section{Landau theory}
We Taylor-expand the energy of a regular hexagonal cell arrangement,
\begin{equation}
    F\left (\xi\right )=\left (\frac{2^{3/2}\cdot 3^{1/4}}{\xi^{1/3}}+2\xi^{2/3}-S_0\right )^2-\frac{\sqrt{2}\cdot3^{1/4}\gamma}{\xi^{1/3}}\>,
\end{equation}
%
%
%
%
in terms of the order parameter $\Xi=\xi-\xi^*$ to obtain
\begin{equation}
    F\approx F_0+a_1\Xi+a_2\Xi^2+a_3\Xi^3+a_4\Xi^4
\end{equation}
with coefficients
\begin{align}
    F_0&=\left (3^{7/6}\cdot 2^{2/3}-S_0\right )^2-2^{2/3}\cdot 3^{1/6}\gamma\>,\\
    a_1&=\frac{2\cdot2^{1/6}}{3\cdot 3^{1/12}}\gamma\>,\\
    a_2&=-\frac{4\cdot 2^{2/3}}{9\cdot 3^{1/3}}\left (3T+\gamma\right )\>,\\
    a_3&=\frac{8\cdot 2^{1/6}}{81\cdot 3^{7/3}}\left (24T+7\gamma\right ),\\
    a_4&=-\frac{4}{729}\left (-54\cdot 6^{1/3}+126\cdot 2^{2/3}\cdot3^{1/6}T+35\cdot 2^{2/3}\cdot 3^{1/6}\gamma\right )\>,
\end{align}
where we defined a generalized temperature as
\begin{equation}
    T=S_0-S_0^*\>.
\end{equation}

\subsection{No lateral tension}
We set $\gamma=0$ and examine the equation of state, ${\rm d}F/{\rm d}\Xi=0$, which reads
\begin{equation}
    \Xi\left [\left (-12\cdot 6^{1/3}+28\cdot 2^{2/3}\cdot 3^{1/6}T\right )\Xi^2-24\cdot 2^{1/6}\cdot 3^{5/12}T\Xi+9\cdot 6^{2/3}T\right ]=0\>.
\end{equation}
Above the critical temperature, i.e., $T>0$,
\begin{equation}
    \Xi_{\pm}=\frac{3\cdot 3^{2/3}T}{2\sqrt{2}\cdot 3^{5/12}T\mp 2^{5/6}\sqrt{T\left (9-5\cdot 2^{1/3}\cdot 3^{5/6}T\right )}}
\end{equation}
Expanding these solutions close to the critical point yields their amplitude
\begin{equation}
    \left | \Xi\right |=3^{2/3}2^{-5/6}T^{1/2}\propto T^{1/2}\>.
\end{equation}

\subsection{Susceptibility}
Again, we start with the equation of state and spell it out for a general case $\left |\gamma\right |>0$:
\begin{equation}
    E=\frac{2\cdot 2^{1/6}\gamma}{3\cdot 3^{1/12}}-\frac{8\cdot 2^{2/3}\left (3T+\gamma\right )}{9\cdot 3^{1/3}}\Xi+\frac{8\cdot 2^{1/6}\left (24T+7\gamma\right )}{27\cdot 3^{7/12}}\Xi^2-\frac{16}{729}\left (-54\cdot 6^{1/3}+126\cdot 2^{2/3}\cdot 3^{1/6}T+35\cdot 2^{2/3}\cdot 3^{1/6}\gamma\right )\Xi^3=0\>.
\end{equation}
Susceptibility is defined as 
\begin{equation}
    \chi=\frac{{\rm d}\Xi}{{\rm d\gamma}}\Big |_{\gamma=0}\>.
\end{equation}
Using implicit differentiation of $E=0$ yields
\begin{equation}
    \frac{\partial E}{\partial\gamma}+\frac{\partial E}{\partial\Xi}\frac{\partial\Xi}{\partial\gamma}=0\>,
\end{equation}
from where it follows
\begin{equation}
    \chi=-\frac{\left (\partial E/\partial\gamma\right )}{\left (\partial E/\partial\Xi\right )}\Big |_{\gamma=0,\Xi=\Xi_{\rm eq}}\>.
\end{equation}
The partial derivatives read
\begin{align}
    \frac{\partial E}{\partial\gamma}&=\frac{2 \cdot 2^{1/6} \,\left [ 81 \cdot 3^{3/4} 
    - 4 \Xi \left ( 27\sqrt{6} - 63\cdot 3^{1/4} \Xi + 70 \sqrt{2} \Xi^2 \right )\right ]}
    {243 \cdot 3^{5/6}}\\
    \frac{\partial E}{\partial\Xi}&=\frac{8}{243} \Big( 
   -9\cdot 6^{2/3} (3T + \gamma)
   + 6\cdot 2^{1/6} 3^{5/12} (24T + 7\gamma)\,\Xi
   - 2 \big( -54\cdot 6^{1/3} + 126\cdot 2^{2/3} 3^{1/6} T 
             + 35\cdot 2^{2/3} 3^{1/6} \gamma \big)\,\Xi^2
\Big)
\end{align}
\underline{For $T<0$}, $\Xi_{\rm eq}=0$ and after inserting $\gamma=0$, we obtain for the susceptibility
\begin{equation}
    \chi_-=-\frac{3^{1/4}}{4\sqrt{2}\left |T\right |}\propto \left |T\right |^{-1}\>.
\end{equation}
Similarly, \underline{for $T>0$}, where $\Xi_{\rm eq}=\pm 3^{2/3}2^{-5/6}T^{1/2}$,
\begin{equation}
    \chi_+=-\frac{3^{1/4}}{8\sqrt{2}T}\propto T^{-1}\>.
\end{equation}
Finally, we note that $\chi_-/\chi_+=2$.

\subsection{Critical isotherm}
For $T=0$ and small $\gamma$, the equation of state simplifies to 
\begin{equation}
    \frac{2^{7/6}}{3^{13/12}}\gamma+\frac{32\cdot 6^{1/3}}{27}\Xi^3=0
\end{equation}
and the critical isotherm reads
\begin{equation}
    \gamma=-\frac{2^{25/6}}{3^{19/12}}\Xi^3\>.
\end{equation}

\section{Cells with variable volumes}
In the case of finite $k_V$, cells can have different in-plane polygon areas even when their heights $h$ are assumed equal. A 2D-projected single-cell energy in this case takes a more general form:
\begin{equation}
    \label{eq_energy_general}
    W_k=k_A\left (A_k-A_0\right )^2+k_p\left (P_k-P_0\right )^2+k_{c}A_kP_k+\omega\>,
\end{equation}
where $k_A=4+h^2k_V$, $A_0=(2S_0+hk_V)/(4+h^2k_V)$, $k_p=h^2$, $P_0=(S_0+\gamma/4)/h$, $k_c=4h$, and $\omega=k_V+S_0^2-(2S_0+hk_V)^2/(4+h^2k_V)-\left (4S_0+\gamma\right )^2/16$, and computing the equilibrium at a given in-plane strain requires minimizing $W=\sum_k W_k$ with respect to $A_k$ and $P_k$, and~$h$. 

To assess the influence of $k_V$, we consider tissues in the floppy regime, where all cells have equal energies, and their shape matches the preferred one. Note that due to the last two terms in~Eq.~(\ref{eq_energy_general}), minimal energy is not necessarily reached at $A_k = A_0$ and $P_k = P_0$. Nevertheless, 
Cell in-plane areas $A_k = A$ are imposed by the in-plane strain and are all equal. Again, this allows us to renormalize in-plane perimeters as $p_k = P_k/\sqrt{A}$. Expressing~Eq.~(\ref{eq_energy_general}) in terms of $p$ and $h$, and minimizing it, yields 
$p = A^{1/2}(S_0 - 2A + \gamma/4)$, exactly matching $p_0$ in the $k_V \to \infty$-case~[Eq.~(5)]. This allows us to conclude that $k_V$ does not effect the rigid-floppy boundary $S_0(\xi, \gamma)$.
\begin{figure*}[htb!]
    \includegraphics[]{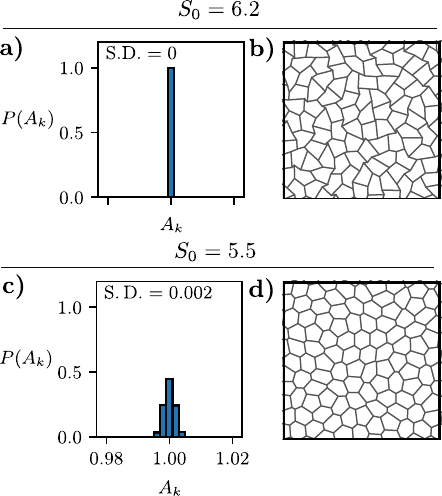}
    \caption{Distributions of in-plane cell areas $A_k$. In compatible states (panel a), all cells assume the preferred in-plane area. For incompatible states (panel b), the distribution broadens slightly, showing  small deviation from the preferred value. In both panels, $\xi=0.93$.}
    \label{FS4}
\end{figure*}

\begin{figure*}[htb!]
    \includegraphics[]{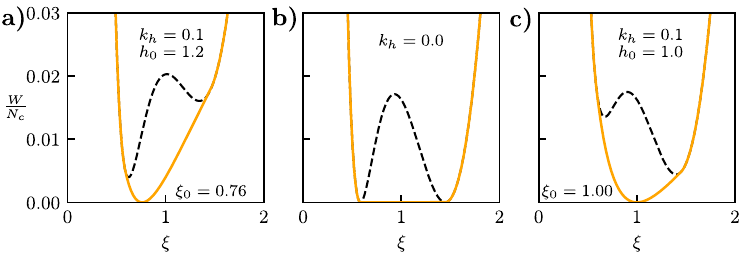}
    \caption{Energy as a function of aspect ratio in a model, where the energy of each cell additionally penalizes deviations of cell height from a preferred height $h_0$ through an energy term $k_h(h_i-h_0)^2$, $k_h$ being the corresponding modulus. Black dashed curves show energies of a tissue, where all cells are regular hexagons. Dashed curves are visible only in the compatible regime, where cells assume irregular rather than regular hexagonal shapes. In the case, where changes of cell height are not penalized (panel b), states for $\xi_-<\xi<\xi_+$ are energy-degenerate, implying a vanishing in-plane bulk modulus. The cell-height energy term lifts this degeneracy as the minimal-energy state is the one, where cells assume the preferred height $h_0$ (panels a and c).}
    \label{FS5}
\end{figure*}

\begin{figure*}[htb!]
    \includegraphics[]{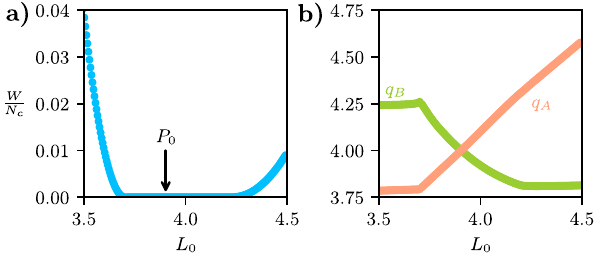}
    \caption{Energy~(a) and in-plane cell shape factors (i.e., area-renormalized perimeters) of the apical and basal sides~(b) as functions of the preferred apical perimeter ($L_0$). Each cell’s energy includes an additional penalty term $k_P(P - L_0)^2$ compared to the original model, where $k_P$ is the modulus and $L_0$ is the preferred apical perimeter. The data correspond to a disordered sample with $d = 0.437$, $\gamma=0$, $\xi=0.93$, and $S_0=6$. (a)~The nonzero energy values identify the incompatible regime, whereas the vanishing energy indicates the compatible states. Here, $P_0$ (marked by the black arrow) denotes the in-plane cell perimeter at which compatibility occurs in the original model. These data show that even with the additional apical-perimeter penalty, a finite region of cell compatibility persists. (b) The corresponding in-plane cell shape factors of the apical and the basal cell sides, $q_A$ and $q_B$, respectively, show that cells are symmetric at $L_0=P_0$, where appical perimeters are not in conflict with $P_0$, and asymmetric otherwise.}
    \label{FS6}
\end{figure*}

\bibliography{apssamp}